\documentclass{article}
\usepackage{amsfonts,amssymb,amsmath}
\usepackage[dvips]{epsfig}
\usepackage[T1]{fontenc}
\usepackage[latin1]{inputenc}
\usepackage{graphicx}
\usepackage[english]{babel}
\usepackage{amsmath}
\usepackage{amssymb}
\usepackage{amsfonts}
\usepackage{color}
\usepackage{rotating}
\usepackage{pdflscape}

\begin{document}

\title{ \bf Surrounded Vaidya Black Holes: \\Apparent Horizon Properties }
\author{Y. Heydarzade$$\thanks{%
email: heydarzade@azaruniv.edu}~~and ~~F. Darabi$$\thanks{%
email: f.darabi@azaruniv.edu} ,
\\{\small Department of Physics, Azarbaijan Shahid Madani University, Tabriz, Iran}}
\maketitle
\begin{abstract}
 We  study the thermodynamical features and
dynamical evolutions of  various apparent horizons associated
with the  Vaidya evaporating black hole  surrounded by the cosmological
fields of dust, radiation, quintessence, cosmological constant-like 
and phantom. In this regard, we address  in detail how do these surrounding fields contribute to the characteristic features of a surrounded dynamical black hole
in comparison to  a dynamical black hole in an empty background. 
\end{abstract}

\section{Introduction}
The Vaidya solution \cite{ahvaidya, ahvaidyajoon1} is one of the non-static solutions of  Einstein
field equations. This solution can be regarded as a dynamical generalization of
the static Schwarzschild black hole solution. This
solution is characterized by a dynamical mass function, i.e $m=m(u)$ where
$u$ is the advanced  time coordinate, and an outgoing null radiation
flow. One can interpret this null radiation  as high frequency electromagnetic or gravitational waves, massless scalar particles or neutrinos.  Based on these properties, this   solution
has been used  for
studying the spherical symmetric gravitational collapse and for exploring
the validity
of the cosmic censorship conjecture \cite{ahccc, ahnaked1, ahnaked2, ahnaked3},  as a model for studying evaporating black holes as well as  Hawking radiation \cite{ahevap1, ahevap3, ahevap4, ahevap6, ahparikh}, among the other
applications. This  solution has been  generalized to the   Vaidya-de Sitter solution \cite{ahmallet1}, Bonnor-Vaidya solution
\cite{ahbonnor},
 Bonnor-Vaidya-de Sitter solution \cite{ahBVdS,
ahPatino, ahMallet*, ahkoberlin, ahsaida}, radiating dyon solution \cite{ahdyon}, and
 the surrounded Vaidya and Bonnor-Vaidya solutions \cite{our1, our2}, see also  \cite{ahgen1, ahgen2} for the other similar generalizations.

 The  connections between black 
holes and  thermodynamics were  initially discovered by Hawking \cite{ahhawking1} and Bekenstein \cite{ahbek}. Black holes 
behave  as  thermodynamic objects  emitting  radiation,  the well-known Hawking
radiation,   from  their  event horizons which is described in the context
of  quantum field theory in curved spacetime. From  thermodynamical point
of view,
there are $(i)$ a characteristic 
temperature for a black hole which is proportional to the surface gravity associated
with its event horizon (EH), and $(ii)$ an entropy 
equal to one quarter of the area of the EH. Interestingly, the apparent  horizon (AH) and the event horizon EH
coincide for  the static or stationary black holes. Then, one cannot distinguish
exactly where is the actual emission surface for the Hawking radiation;   AH or   EH?  For the case of  Vaidya
black hole, it is found in \cite{ahhajicek}
 that the  Hawking radiation originates always in regions having geometry  near to the geometry in a close
neighborhood of AH. Then, EH 
appears as radiating surface only if it passes through such regions, similar
to what happens for the case of static Schwarzschild black hole.
 Considering the collapse of a spherical shell leading to a black hole with
 the Vaidya solution \cite{ahhiscock},
 it is suggested  that in order to maintain a local thermodynamics,  the entropy of the black hole should be identified as $\mathcal{A}_{AH}/4$ where $\mathcal{A}_{AH}$ is the area of the black hole's AH.
Also, the fact that thermodynamics of a dynamical
black hole can be built successfully on the AH is verified in \cite{ahliou, ahzhou}.

In this paper, we aim to study the  properties of the various horizons, mainly the AH, associated
with the  Vaidya evaporating black hole solution surrounded by the cosmological
fields of dust, radiation, quintessence, cosmological constant-like
and phantom.  In this regard, we explore the surrounded Vaidya solutions
introduced in
\cite{our1} using the approach  used in \cite{ahyork1, ahmallet}.
The organization of the paper is as follows. In section 2, we review briefly
the
surrounded Vaidya black hole solutions introduced in  \cite{our1}. In section 3, we discuss on the 
proprieties of these solutions in a general form. In section 4, we study
in detail the properties of these solutions for the surrounding fields of
dust, radiation, quintessence, cosmological constant-like and phantom. In section 5, we give our concluding remarks
with introducing  our two new works.

\section{The Surrounded Evaporating Vaidya Black Holes}
The   Vaidya black hole surrounded by generic fields in the advanced
time coordinate $u$ is obtained  in \cite{our1} as 
 \begin{equation}\label{Vaidya}
dS^{2} =-\left(1-\frac{2M(u)}{r} -\frac{N_s(u)}{r^{3\omega_s+1}} \right)du^2
+2 dudr+r^2
d\Omega_{2}^2,
 \end{equation}
where $M(u)$, $N_s(u)$ and $\omega_s$  are the  time dependant
 decreasing mass of the evaporating Vaidya black hole,
the normalization parameter   and the equation of state parameter of the surrounding  field, respectively. The corresponding total energy-momentum tensor ${T^{\mu}}_{\nu}$ is given by \cite{our1, our2, Kiselev, our3} 
\begin{equation}\label{t**}
{T^{\mu}}_{\nu}={\tau^{\mu}}_{\nu}+{\mathcal{T}^{\mu}}_{\nu},
\end{equation}
where ${\tau^{\mu}}_\nu$  is the energy-momentum tensor
associated to the Vaidya null radiation as%
\begin{equation}\label{null}
{\tau^{\mu}}_{\nu}=\sigma k^{\mu}k_{\nu}, 
\end{equation}
such that  $\sigma=\sigma(u,r)$ is the density of the ``outgoing  radiation 
flow'' and  $k^\mu$ is a null vector field
directed radially outward,  and    ${\mathcal{T}^{\mu}}_{\nu}$ is the energy-momentum
tensor of the surrounding dynamical field as 
\begin{eqnarray}\label{lala*}
&&{\mathcal{T}^{0}}_{0}={\mathcal{T}^{1}}_{1}=-\rho_s(u,r),\nonumber\\
&&{\mathcal{T}^{2}}_{2}={\mathcal{T}^{3}}_{3}=\frac{1}{2}\left(1+3\omega_s\right)\rho_s(u,r),
\end{eqnarray}
where the energy density $\rho_s(u,r)$ of the
surrounding field is given by
\begin{equation}\label{density}
\rho_s(u,r)=-\frac{3\omega_s N_s(u)}{r^{3(\omega_s+1)}}.
\end{equation}
The  weak energy condition on the energy density (\ref{density}) of the surrounding
field requires
\begin{equation}\label{WEC**}
\omega_s N_{s}(u)\leq0,
\end{equation}
implying that for the surrounding fields with $\omega_s\geq0$ and  $\omega_s\leq0$, we need $N_{s}(u)\leq0$ and $N_{s}(u)\geq0$, respectively. Regarding (\ref{Vaidya}) and (\ref{t**}), the original Vaidya evaporating
black hole solution can be recovered by turning off
the background field, i.e setting $\rho_s(u,r)=0$.

Following  \cite{ahyork1,Carter, Poisson}, we can rewrite the metric of our surrounded Vaidya black hole spacetime (\ref{Vaidya}), using the  null-vector  decomposition, in the following  form 
\begin{equation}
g_{\mu\nu}=-n_{\mu}l_{\nu}-l_{\mu}n_{\nu}+h_{\mu\nu},
\end{equation}
where $n_\mu$ and $l_\mu$ are ingoing and outgoing null vectors, respectively,
given by
\begin{eqnarray}\label{conditions}
n_{\mu}&=&-\delta^{u}_{\mu},\nonumber\\
l_{\mu}&=&-\frac{1}{2}\left(1-\frac{2M(u)}{r} -\frac{N_s(u)}{r^{3\omega_s+1}} \right)\delta^{u}_{\mu}+\delta^{r}_{\mu},
\end{eqnarray}
 and
\begin{equation}
h_{\mu\nu}=r^2\delta^{\theta}_{\mu}\delta^{\theta}_{\mu}+r^2 sin^2\theta\delta^{\phi}_{\mu}\delta^{\phi}_{\mu},
\end{equation}
is the two dimensional metric for $u=constant$ and $r=constant$ surfaces such that  $n_\mu$,
$l_\mu$ and $h_{\mu\nu}$ satisfy the following normalization conditions
\begin{eqnarray}\label{normalization}
&&n_{\mu}n^{\mu}=l_{\mu}l^{\mu}=0,\nonumber\\
&&l_{\mu}n^{\mu}=-1,\nonumber\\
&&l^{\mu}h_{\mu\nu}=n^{\mu}h_{\mu\nu}=0.
\end{eqnarray}
As we will see in the following section, the null vector decomposition of the
spacetime metric is more appropriate for studying its various geometric properties.
%
\section{The  Location and Properties of the Horizons}
It is a well known fact that the black holes in general have not  single horizons to  characterize completely their local and global structures, see \cite{ahyork1,ahNielsen} as instances.  They possess three important loci: $(i)$
the future EH which is a null three-surface describing the locus of outgoing future-directed null 
geodesic rays that never  reach arbitrarily large distances from the black hole, $(ii)$
the AH which is the outermost marginally trapped three-surface for the 
outgoing photons.  Classically, the AH can be either null or spacelike surfaces, and
consequently it can move causally or non-causally. Considering the black hole quantum radiance, the AH 
can also be a timelike surface, and $(iii)$  the timelike limit 
surface (TLS) which is a  horizon-like locus. For the black holes possessing
an exact timelike 
Killing vector field $\partial/\partial t$ in their exterior regions,    TLS is the surface given by $g(\partial_t, \partial_t)=g_{tt}=0$. For black holes with a small 
dimensionless accretion or luminosity $L=-\frac{dM}{du}=-\dot M(u)$, one can define the TLS (or quasi-static limit) as the locus where $g(\partial_u, \partial_u)=g_{uu}= 0$,  $\partial/\partial u$ being the timelike vector field for  an 
observer sitting at rest at the far distance from the black hole. Generally, the TLS can be a null, timelike, 
or a spacelike three-surface. We note that 
when $L$ refers to the quantum radiance, i.e given by the Hawking relation $L = L_H \propto\hslash /M^2$, it is usually very small. In this case, it represents a black hole radiating an small amount of quanta  in 
any reasonable time interval. However, the fluctuations in the radiation
 could be very large, at least 
comparable to $L$ itself. Thus, we consider $L$   as an average value
such that the terms with $\mathcal{O}(L^2)$ are negligible.    We also mention that for the spherical symmetric solutions of
$R_{\mu\nu} = 0$,  there is a 
degeneracy for the horizons which means that  all these three horizons coincide at  $r=2M$, known as the Schwarzschild radius. In general, such a coincidence
of various horizons happens for the stationary situations. However, 
in the case of  small accretion or luminosity, i.e for $R_{\mu\nu}= \mathcal{O}(L)$,  this degeneracy is 
partially lifted even for the  spherical symmetric spacetimes. For the case
of $R_{\mu\nu}= \mathcal{O}(L)$,   we have  $AH=TLS$ but $EH\neq AH$.  For the case of  $R_{\mu\nu}=0$ and without spherical symmetry but preserving stationarity, as for a Kerr hole, we have $EH=AH$ but $EH\neq TLS$. So, we see that these 
three horizons do not coincide generally and they are very sensitive to small perturbations. Moreover, by including
the generic background fields  in our solution (\ref{Vaidya}), we will consider $\dot N_s(u)$  here also as an average value and the terms with $\mathcal{O}(\dot N_s^2)$ are negligible.

In general, the structure and dynamics of various horizons for dynamical spacetimes can
be understood using the    behavior  of  null geodesic  congruences, 
governed by the Raychaudhuri equation \cite{Poisson}
\begin{equation}\label{ray}
\frac{d\Theta}{du}=\mathcal{K}\Theta-\frac{1}{2}\Theta^2-\sigma^2 +\omega^2-R_{\mu\nu}l^{\mu}l^{\nu},
\end{equation}
where $\mathcal{K}$, $\Theta$, $\sigma^2=\sigma_{\mu\nu}\sigma^{\mu\nu}$, $\omega^{2}=\omega_{\mu\nu}\omega^{\mu\nu}$ and $R_{\mu\nu}$ are the
surface gravity, expansion scalar,  shear scalar,
vorticity scalar,  and Ricci tensor, respectively.    The expansion scalar
for the null rays parameterized 
by the parameter $u$ is given by \cite{ahyork1, ahFaraoni}
\begin{equation}\label{theta}
\Theta=\nabla_{\mu}l^{\mu}-\mathcal{K},
\end{equation}
where  $\nabla_{\mu}$  is  the  covariant  derivative  operator corresponding
to the metric (\ref{Vaidya}). The  surface gravity is given by
the non-affinely parameterized geodesic equation with the null tangent vector
$l^{\mu}$ as  \cite{ ahFaraoni, ahNielsen, ahNielsen1, ahNielsen2}
\begin{equation}\label{surf}
l^{\mu}\nabla_{\mu}l^{\nu}=\mathcal{K}l^{\nu}.
\end{equation}
Using the normalization conditions (\ref{normalization}) on the null vectors $n^{\mu}$ and $l^{\mu}$,  one  can obtain the surface gravity  in (\ref{surf}) in the following form
\begin{equation}\label{SG}
\mathcal{K}=-n^{\nu}l^{\mu}\nabla_{\mu}l_{\nu}.
\end{equation}
Spherically symmetric non-rotational spacetimes, such as our metric  (\ref{Vaidya}), are vorticity and shear free, i.e $\sigma_{\mu\nu}=0=\omega_{\mu\nu}$. Then, for these spacetimes, the location, structure
and dynamics of the various horizons depend  on the surface gravity $\mathcal{K}$,
the expansion scalar   $\Theta$ and the Ricci tensor $R_{\mu\nu}$.  One can follow York 
\cite{ahyork1} and obtain the  location of the AH and EH for a dynamical radiating
black hole up to  $\mathcal{O}(L)$ as  \begin{itemize}
 \item AH is 
defined at the surface  such  that $\Theta\simeq 0$,
\item  EH is the null 
surface such  that $\frac{d\Theta}{du} \simeq0$.
\end{itemize}
Thus, one realizes that the  EH  can be obtained using the Raychaudhuri equation (\ref{ray}), and the   AH  can  be 
determined  by  the equation (\ref{theta}).

Using the equations  (\ref{conditions})  and (\ref{SG}), one can obtain the  general dynamical surface gravity associated to our metric (\ref{Vaidya}) as 
\begin{equation}\label{sur*}
\mathcal{K}=\frac{1}{2}\left(\frac{2M(u)}{r^{2}} +\frac{(3\omega_s +1)N_s(u)}{r^{3\omega_s
+2}}\right).
\end{equation}
Then, by using the equations (\ref{conditions}),  (\ref{theta}),  and (\ref{sur*}), the expansion scalar can be obtained as 
\begin{equation}\label{exp1}
\Theta=\frac{1}{r}\left(1-\frac{2M(u)}{r} -\frac{N_s(u)}{r^{3\omega_s+1}}\right).
\end{equation} 
Regarding (\ref{Vaidya}) and (\ref{exp1}), we see that the solution to $g(\partial_u, \partial_u)=g_{uu}= 0$ representing
the TLS is the same as the solution to $\Theta=0$ representing
the AH location except the trivial solution $r=\infty$ for the latter. Then, the AH and the TLS  coincide for our spacetime (\ref{Vaidya}). 

The Hawking temperature on the horizon  (apparent or event) of black hole can be determined using its corresponding surface gravity as
\begin{equation}
\mathcal{T}_{H}=\frac{\mathcal{K}}{2\pi}.
\end{equation}
Then, regarding (\ref{sur*}), we see that temperature is also a dynamical
quantity with respect to the time coordinate $u$.
Also, the semiclassical Bekenstein-Hawking entropy at the horizon is given
by \cite{ahbek}
\begin{equation}
\mathcal{S}_H=\frac{\mathcal{A}_H}{4},
\end{equation} 
where $\mathcal{A}_H$ is the area of the horizon given by
\begin{equation}
\mathcal{A}_H=\int_{0}^{2\pi}\int_{0}^{\pi}\sqrt {g_{\theta\theta}g_{\phi\phi}}d\theta
d\phi=4\pi r_{H}^2.
\end{equation} 
In the following of this section, we show that the condition $\frac{d\Theta}{du} \simeq0$ determining
 the location of the EH up to $\mathcal{O}(L)$ is equivalent to the requirement that 
the acceleration of the congruences of null geodesics  vanishes at the EH,
i.e 
\begin{equation}\label{acceleration}
\ddot r_{EH}(u)=\frac{d^2 r}{du^2}\mid_{r=r_{EH}} \simeq0.
\end{equation}
To prove this, using the metric (\ref{Vaidya}), for a radial null geodesic
parametrized by the parameter $u$, we have
\begin{equation}\label{velocity}
\dot r(u)=\frac{dr}{du}=\frac{1}{2}\left[1-\frac{2M(u)}{r} -\frac{N_s(u)}{r^{3\omega_s+1}}\right],
\end{equation}
and consequently, we find
\begin{equation}\label{ddot}
\ddot r(u)=\frac{1}{2}\left[\frac{2L}{r}-\frac{\dot N_s(u)}{r^{3\omega_s+1}}+\dot r(u)\left(\frac{2M(u)}{r^2} +\frac{(3\omega_s+1)N_s(u) }{r^{3\omega_s+2}}\right)\right].
\end{equation}
Then, using the equations (\ref{sur*}), (\ref{exp1}) and (\ref{ddot}), we can rewrite
the equation (\ref{acceleration}) as
\begin{equation}\label{surthet}
\mathcal{K}\Theta\mid_{r=r_{EH}} =-\frac{2L}{r_{EH}^2}+\frac{\dot N_s(u)}{r_{EH}^{3\omega_s+2}}.
\end{equation}
On the other hand, using the metric (\ref{Vaidya}) and the null vector $l^\mu$
defined in (\ref{conditions}), we find 
\begin{equation}\label{R}
R_{\mu\nu}l^{\mu}l^{\nu}\mid_{r=r_{EH}} =-\frac{2L}{r_{EH}^2}+\frac{\dot N_s(u)}{r_{EH}^{3\omega_s+2}}.
\end{equation}
Thus, by using the equations (\ref{surthet})
and (\ref{R}) and neglecting $\Theta^2=\mathcal{O}(L^2, \dot N^2_s, L\dot
N_s)$ in the Raychaudhuri equation (\ref{ray}), we arrive
at 
\begin{equation}\label{EH}
\frac{d\Theta}{du}\mid_{r_{EH}}\simeq0.
\end{equation}
Hence, the condition (\ref{EH}) determining
 the location of the event horizon becomes equivalent to solving the equation (\ref{surthet})
in $\mathcal{O}(L , \dot N_s)$ resulting from the vanishing acceleration of the congruences of null geodesics  at EH. Then, the discussion in \cite{ahyork1}
for the EH location is also valid and verified here for the surrounded evaporating
black holes.

As mentioned
previously, in contrast to the stationary spacetimes, the local definitions of the various horizons  do not necessarily coincide
with the location of the EH for dynamical black holes \cite{ahNielsen}. For such dynamical spacetimes,  one is left with the question: ``\textit{For
which surface should one define the black hole area, surface gravity, temperature
or entropy?}''
The canonical choice is to use the EH. However, as mentioned in
the introduction, there are 
some evidences that it is the AH, and not the EH, that plays the key
role in the Hawking radiation \cite{ahhajicek, ahhiscock, ahliou, ahzhou}, see
also \cite{ahVisser, ahHayward, ahAshtekar}. This finding has became a key point in hopes to demonstrate
the Hawking radiation in the laboratory using the models of analogue gravity \cite{ahLiberati}. Thus, regarding the importance of AH, we first introduce some specific subclasses of our general surrounded Vaidya black hole solution
(\ref{Vaidya}) and then, we investigate in detail the AH properties of these solutions including the AH location and dynamics as well as their associated Hawking temperature
and Bekenstein-Hawking entropy. We consider the detailed study of  EHs as
our next work.
\section{Apparent Horizon Properties for the Evaporating Vaidya Black Hole in Generic Backgrounds}
\subsection{Evaporating Vaidya Black Hole in an Empty Background} 
In this section, we first explore the AH properties of the Vaidya black
hole in empty background. This provides us with the possibility of
comparing the properties of this black hole in the various backgrounds with its
original form in an empty space.

The case of Vaidya black hole in an empty space corresponds to $N_s(u)=0$. Then, the  metric (\ref{Vaidya}) takes the
original Vaidya  solution form as
\begin{equation}\label{01*}
dS^{2} =-\left(1-\frac{2M(u)}{r}  \right)du^2
+2dudr+r^2
d\Omega_{2}^2.
 \end{equation}
The  surface gravity associated to the metric (\ref{01*}) is given by 
\begin{equation}\label{surfempty}
\mathcal{K}=\frac{M(u)}{r^{2}} ,
\end{equation}
where the expansion scalar for the outgoing null geodesics reads as
\begin{equation}\label{expempty}
\Theta=\frac{1}{r}\left(1-\frac{2M(u)}{r} \right).
\end{equation}
Neglecting the trivial solution of $\Theta=0$ at $r=\infty$, the location
of  the AH is given by  
\begin{equation}\label{ahempty}
r_{AH}(u)=2M(u). 
\end{equation}
Then, the dynamics of the single AH (\ref{ahempty}) is governed by the relation 
\begin{equation}\label{ahdotempty}
\dot r_{AH}(u)=-2 L.
\end{equation}
This shows that the single AH is moving inward during the evaporation, i.e
it is shrinking.
 For this case, the nature of the AH (\ref{ahempty}) can be determined by
looking at the induced metric on the horizon as
\begin{equation}\label{0empty}
dS_{{AH}}^2 =-4\,L\,du^2+4M^2(u)
d\Omega_{2}^2,
 \end{equation}
showing that AH is a timelike surface for an evaporating BH.   
 The Hawking temperature on the AH (\ref{ahempty}) can be determined using the surface gravity as
\begin{equation}\label{tahempty}
\mathcal{T}_{AH}=\frac{1}{2\pi}\frac{M(u)}{ r_{AH}^2}=\frac{1}{8\pi}\frac{1}{M(u)}.
\end{equation}
This relation implies that the  temperature  associated with the AH (\ref{ahempty}) rises up
by decreasing the mass of  black hole during the evaporation. Finally, the semiclassical Bekenstein-Hawking entropy at AH (\ref{ahempty}) is given
by
\begin{equation}\label{sempty}
\mathcal{S}_{AH}=\pi r_{AH}^2=4\pi M^2(u).
\end{equation}

\subsection{ Evaporating Vaidya Black Hole Surrounded by  the Dust Field} 

For the case of Vaidya black hole surrounded by the dust field with the equation
of state parameter $\omega_d=0$ \cite{Kiselev, Vik}, the metric (\ref{Vaidya}) takes the
following form
\begin{equation}\label{011}
dS^{2} =-\left(1-\frac{2M(u)}{r} -\frac{N_{d}(u)}{r} \right)du^2
+2dudr+r^2
d\Omega_{2}^2,
 \end{equation}
 where $N_{d}(u)$ is the normalization parameter for the dust field surrounding
the back hole, with
the dimension of $\left[N_d \right]=l$ where $l$ denotes the length. Regarding
(\ref{011}), the black hole in the dust background appears  as
a black hole with an effective
dynamical mass $2M_{eff}(u)=2M(u)+N_d(u)$. In the following, we see that
how the presence of this effective mass    modifies the thermodynamics,
causal structure
and Penrose diagrams of the original  Vaidya black hole. 

The  surface gravity associated to the metric (\ref{011}) is given by 
\begin{equation}\label{sur}
\mathcal{K}=\frac{1}{2}\left(\frac{2M(u)}{r^{2}} +\frac{N_{d}(u)}{r^{2}}\right).
\end{equation}
We see that the dust background appears with a positive surface
gravity which increases the gravitational attraction of the black hole. The expansion scalar for the outgoing null geodesics reads as 
\begin{equation}\label{exp*}
\Theta=\frac{1}{r}\left(1-\frac{2M(u)}{r} -\frac{N_{d}(u)}{r}\right).
\end{equation}
 Then, we find the  solution to  $\Theta=0$ determining the location
of the single physical AH
as 
\begin{equation}\label{ahd}
r_{AH}(u)=2M_{eff}(u),
\end{equation}
where $r_{AH}(u)$ corresponds to a modified larger AH compared with  the case of black hole in an
empty space given by (\ref{ahempty}). Then, the surrounding background dust
field contributes
positively to the black hole mass and size of AH . The dynamics of the AH (\ref{ahd})
is governed by the relation
\begin{equation}\label{mimi}
\dot r_{AH}(u)=2\dot
M_{eff}(u)=-2L+\dot N_{d}(u).  
\end{equation}
Then, the AH behaviours for the cases of evaporating
black hole in a static and dynamical dust backgrounds are different. For this case, we
have the following possibilities: 
\begin{itemize}
\item For $\dot N_d (u)=0$, representing the static surrounding dust field, the AH (\ref{ahd}) is shrinking. Regarding (\ref{ahempty}), this case also shows that the dynamics of AH  for an evaporating black hole surrounded by a static  dust
field  is the same as the black hole in the empty space.  
\item For $\dot N_d (u)=2L$, the AH (\ref{ahd})
is frozen. This case shows that the black hole input by the surrounding field is equal to its output radiation.

\item For $-2L+\dot N_{d}(u)>0$, the AH (\ref{ahd}) is expanding. This case requires
$\dot N_{d}(u)>2L>0$ which shows an
evaporating black hole in a collapsing dust field. 
\item For $-2L+\dot N_{d}(u)<0$, the AH (\ref{ahd})
is shrinking. This case requires $0<\dot N_{d}(u)<2L$ or
$\dot N_{d}(u)<0$ showing an
evaporating black hole in a slowly collapsing or a decaying dust field, respectively. The AH (\ref{ahd}) with $0<\dot N_{d}(u)<2L$ or
$\dot N_{d}(u)<0$, shrinks slower or faster than  the AH  of black hole in a
static
dust background.\end{itemize}
 The nature of  AH (\ref{ahd}) can be determined by
looking at the induced metric on it as

\begin{equation}\label{0d}
dS_{AH}^2 =2\left(-2L+\dot N_{d}(u)\right)du^2+4M_{eff}^2
d\Omega_{2}^2,
 \end{equation}
which implies that the AH (\ref{ahd}) is a timelike surface for an
evaporating black hole in the static dust field and for a dust field with $-2L+\dot N_{d}(u)<0$, whereas it
is a spacelike surface for a dust field with $-2L+\dot N_{d}(u)>0$.
For the specific case of $\dot N_d (u)=2L$, AH (\ref{ahd}) is a null surface.

 The Hawking temperature of the AH (\ref{ahd}) of black hole is given by
 \begin{equation}\label{tahd}
\mathcal{T}_{AH}=\frac{1}{2\pi}\frac{M_{eff}(u)}{ r_{AH}^2}=\frac{1}{8\pi}\frac{1}{M_{eff}(u)}.
\end{equation}
Then, we observe that the  temperature associated with the
AH  (\ref{ahd}) for the black hole in dust field is lower than the temperature of AH (\ref{ahempty}) of the
black hole in an empty space.
Finally, the semiclassical Bekenstein-Hawking entropy at the AH (\ref{ahd}) is given
by
\begin{equation}
\mathcal{S}_{AH}=\pi r_{AH}^2=4\pi M_{eff}^2(u),
\end{equation}
which implies that the entropy associated with the AH (\ref{ahd}) in a dust field background is higher than the case of the black hole in an empty space.

\subsection{ Evaporating Vaidya Black Hole Surrounded by the Radiation Field} 
For the case of Vaidya black hole surrounded   by a radiation field with
the equation of state parameter  $\omega_r=\frac{1}{3}$ \cite{Kiselev, Vik}, the metric (\ref{Vaidya}) takes the
following form 
\begin{equation}
dS^{2} =-\left(1-\frac{2M(u)}{r} -\frac{N_{r}(u)}{r^{2}} \right)du^2
+2dudr+r^2
d\Omega_{2}^2,
 \end{equation}
where $N_{r}(u)$ denotes the normalization parameter for the radiation field
surrounding the black hole, with
the dimension of $[N_r ]=l^2$.  Regarding the positive energy condition on the surrounding radiation field, represented by the relation
(\ref{WEC**}), it is required that $N_r(u)<0$. Then, by defining the positive parameter $\mathcal{N}_r(u)=-N_r(u)$, we arrive at the following metric
\begin{equation}\label{1/3}
dS^{2}=-\left(1-\frac{2M(u)}{r}+\frac{\mathcal{N}_{r}(u)}{r^2}\right)du^2+2
dudr+r^2d\Omega^2.
\end{equation}
This metric looks like a radiating charged
Vaidya black hole with the dynamical charge $Q^{2}(u)=\mathcal{N}_r(u)$. Then, this result can be
 interpreted as the positive contribution of the characteristic feature
of the surrounding radiation field  as the effective charge-like term for the Vaidya
black hole, see the Bonnor-Vaidya solution \cite{ahbonnor}.  The appearance of such a charge-like term in the metric
changes the causal structure and Penrose diagrams of this  solution in comparison to the original Vaidya black hole.
A similar effect in the causal structure of spacetime happens when one adds charge to the static Schwarzschild black hole leading to Reissner-Nordström black hole.
Then, turning off the background radiation field which surrounds the dynamical
Vaidya
black
hole  is equal to turning off the charge
in the static Reissner-Nordström case. 

The  surface gravity associated to the metric (\ref{1/3}) is given by
\begin{equation}\label{surr}
\mathcal{K}=\frac{1}{2}\left(\frac{2M(u)}{r^{2}} -\frac{2\mathcal{N}_{r}(u)}{r^{3}}\right).
\end{equation}
Then, we see that the radiation background appears with a negative surface
gravity which decreases the gravitational attraction of the black hole, exactly
the same as the charge effect in Reissner-Nordström black hole.
The expansion scalar for the outgoing null geodesics can be obtained as 
\begin{equation}\label{expr}
\Theta=\frac{1}{r}\left(1-\frac{2M(u)}{r} +\frac{\mathcal{N}_{r}(u)}{r^{2}}\right).
\end{equation}
We can find the  solutions to  $\Theta=0$ determining the locations
of the AH
as  
\begin{equation}\label{rr-}
r_{1}(u) =M(u)-\sqrt{M^2(u)-\mathcal{N}_{r}(u)},
\end{equation}
and 
\begin{equation}\label{rr+}
r_{2}(u) = M(u)+\sqrt{M^2(u)-\mathcal{N}_{r}(u)}.
\end{equation}
Then, depending on the sign of the discriminant $\Delta(u) = M^{2} (u)- \mathcal{N}_{r}(u)$, we have the following three
situations.
\begin{itemize}
\item For $\Delta(u)<0$, we have no real solutions. Then, there is no physical AH.
\item For the extremal case of $\Delta(u)=0$, equivalent to $\mathcal{N}_{r}(u)=M^2(u)$,
the solutions (\ref{rr-}) and (\ref{rr+}) coincide with a single extremal AH, i.e $r_{1}(u) =r_{2}(u)=M(u)$. 
\item For $\Delta(u)>0$, there are two real physical AHs as $r_{AH^{-}}(u)=r_{1}(u)$
and $r_{AH^{+}}(u)=r_2(u)$ for the black hole in the radiation background in contrast to the black hole in the empty
space and in the dust field background having a single AH.
\end{itemize}

For the extremal case of $\Delta(u)=0$, there is only one single  extremal
AH given by 
\begin{equation}\label{extr}
r_{AH^{ext}}(u)= r_{1}(u) =r_{2}(u)=M(u), 
\end{equation} which is  half of the dynamical Schwarzschild radius. Then, the dynamics of the single extremal AH (\ref{extr}) is governed by the relation 
\begin{equation}\label{ah*}
\dot r_{AH^{ext}}(u)=- L.
\end{equation}
This shows that the single extremal AH (\ref{extr}) for the black hole in the radiation
background  is moving inward during the evaporation, i.e
it is shrinking.
 For this case, the nature of  extremal AH (\ref{extr}) can be determined by
looking at the induced metric on it as
\begin{equation}\label{0r}
dS_{{AH}^{ext}}^2 =-2L\,du^2+M^2(u)
d\Omega_{2}^2,
 \end{equation}
showing that the extremal AH (\ref{extr}) is a timelike surface for an evaporating BH in the radiation background.   
 The Hawking temperature on the extremal AH (\ref{extr})  can be determined using the surface gravity as
\begin{equation}\label{tahextr*}
\mathcal{T}_{AH^{ext}}=\frac{1}{4\pi}\left(\frac{2M(u)}{ r_{AH^{ext}}^2}-\frac{2M^2(u)}{ r_{AH^{ext}}^3}\right)=0,
\end{equation}
showing that the extremal AH (\ref{extr}) is a zero temperature surface.
 One may interpret this zero temperature as the suppressed or zero Hawking radiation for
 the extremal black hole in the radiation background.
 Finally, the semiclassical Bekenstein-Hawking entropy at AH (\ref{extr}) is given
by
\begin{equation}
\mathcal{S}_{AH^{ext}}=\pi r_{AH^{ext}}^2=\pi M^2(u).
\end{equation}
Then, the entropy of the extremal AH (\ref{extr})  is a quarter of the entropy associated to the single AH of the black hole in an empty space.

For the case of $\Delta(u)>0$, the location of the AHs (\ref{rr-}) and (\ref{rr+})
for the dilute radiation background,
i.e. $\mathcal{N}_r(u)\ll M(u)$,  are given by
\begin{equation}\label{lol}
r_{AH^-}(u)=  \frac{\mathcal{N}_r(u)}{2M(u)}+\mathcal{O}(\mathcal{N}_r^2),
\end{equation}
and
\begin{equation}\label{hol}
r_{AH^+}(u)= 2M(u)-\frac{\mathcal{N}_r(u)}{2M(u)}+\mathcal{O}(\mathcal{N}_r^2).
\end{equation}
 It is seen that the contribution of the surrounding background radiation field leads
to ($i$) the existence of a small inner AH (\ref{lol}) vanishing with $\mathcal{N}_r(u)\rightarrow0$, and ($ii$) an smaller outer AH (\ref{hol}) in comparison to the case of the AH (\ref{ahempty}) for the black
hole in an empty background in which at the limit of  $\mathcal{N}_r(u)\rightarrow0$  we recover the dynamical Schwarzschild radius $r(u)=2M(u)$. The dynamics of the AHs (\ref{lol}) and (\ref{hol}) are given by 
\begin{equation}\label{ahr-}
\dot r_{AH^{-}}(u)= \frac{L\,\mathcal{N}_r(u)}{2M^{2}(u)}+\frac{ \mathcal{\dot
N}_r(u)}{2M(u)}+\mathcal{O}(\mathcal{N}_r^2),
\end{equation}
and
\begin{equation}\label{ahr+}
\dot r_{AH^{+}}(u)= -2L-\frac{L\,\mathcal{N}_r(u)}{2M^2(u)}
-\frac{ \mathcal{\dot N}_r(u)}{2M(u)}+\mathcal{O}(\mathcal{N}_r^2).
\end{equation}
Then, the AH behaviours for the cases of evaporating
black hole in a static and dynamical radiation backgrounds are different. For this case, we have the following possibilities: 
\begin{itemize}
\item For $\mathcal{\dot N}_r(u)=0$, representing the static surrounding radiation field,  the inner AH (\ref{lol})  is moving outward (is expanding)  while the outer AH (\ref{hol}) is moving inward (is shrinking) during the evaporation. By stopping the evaporation, if possible, both of the horizons will be frozen, i.e $\dot r_{AH^{-}}(u)=\dot
r_{AH^{+}}(u)=0$.
Another interesting result is that
regarding (\ref{ahempty}) and (\ref{ahr+}), one realizes that the outer AH of the black hole  in  the static radiation field (\ref{hol})
is shrinking faster than the case of black hole in an empty background.  
\item For $\mathcal{\dot N}_{r}(u)>0$,  the inner and outer AHs (\ref{lol})
and (\ref{hol}) are moving outward  and inward, respectively,   but both  moving faster than  the AH
of the black hole in the static radiation background. This case represents an
evaporating black hole in a collapsing radiation field.
\item For $\mathcal{\dot N}_{r}(u)<0$, it is possible that the inner and outer AHs (\ref{lol}) and (\ref{hol})  move again outward  and inward, respectively,   but both  will
be slower than  the AH
of the black hole in the  radiation backgrounds with $\mathcal{\dot N}_r(u)=0$
and $\mathcal{\dot N}_{r}(u)>0$. In the particular case of $L\,\mathcal{N}_r(u)=
-\mathcal{\dot N}_r(u) M(u)$, the inner AH (\ref{lol})
 will be frozen while the outer AH (\ref{hol}) will
move inward exactly the same as  AH (\ref{ahempty}) of the black hole in an
empty background. Reversely, if the term $\frac{ \mathcal{\dot
N}_r(u)}{2M(u)}$ overcomes  the other terms in (\ref{ahr-}) and (\ref{ahr+}),
the inner and outer AHs (\ref{lol}) and (\ref{hol})  will move inward and outward, respectively.
\end{itemize}

 The nature of  AHs (\ref{lol}) and (\ref{hol}) can be determined by
looking at the induced metrics on them up to $\mathcal{O}(\mathcal{N}_r)$ as
\begin{equation}\label{dsahr-}
{dS^{2}_{AH^-}}=2\left( \frac{L\,\mathcal{N}_r}{2M^{2}(u)}+\frac{ \mathcal{\dot
N}_r(u)}{2M(u)}\right) du^2+
r_{AH^{-}}^2 d\Omega_{2}^2,
\end{equation}
and
\begin{equation}
dS^{2}_{AH^+}=2\left(  -2L-\frac{L\,\mathcal{N}_r}{2M^2(u)}-\frac{ \mathcal{\dot
N}_r(u)}{2M(u)} \right) du^2+ r_{AH^{+}}^2d\Omega_{2}^2.
 \end{equation}
Then, for $\mathcal{\dot N}_r(u)\geq0$ the inner and outer AHs (\ref{lol}) and (\ref{hol})  are spacelike
and timelike respectively while for $\mathcal{\dot N}_r(u)<0$, we see that
the natures of AHs (\ref{lol}) and (\ref{hol})   depend on the value of $\frac{ \mathcal{\dot
N}_r(u)}{2M(u)}$ term relative to the other terms in the above parentheses. For
the totally positive parentheses, AHs (\ref{lol}) and (\ref{hol})   are spacelike whereas they are timelike for the totally negative parentheses . For the specific case of $L\,\mathcal{N}_r(u)=
-\mathcal{\dot N}_r(u) M(u)$, the inner AH (\ref{lol}) will be null, whereas the outer AH (\ref{hol})  will be timelike.

The Hawking temperature on the AHs (\ref{lol}) and (\ref{hol}) of the black hole can be determined as
\begin{eqnarray}\label{tahr-}
\mathcal{T}_{AH^-}&=&\frac{1}{2\pi}\left(\frac{M(u)}{(M(u)-\sqrt{M^2(u)-\mathcal{N}_r(u)})^{2}} -\frac{\mathcal{N}_r(u)}{(M(u)-\sqrt{M^2(u)-\mathcal{N}_r(u)})^{3}}\right)\nonumber\\&=&\frac{1}{2\pi}\left(
-4\frac{{M}^{3}(u)}{{\mathcal{N}^{2}_r}(u)}+4\frac{M(u)}{{\mathcal{N}_r(u)}}-\frac{1}{4M(u)}\right)+\mathcal{O}(\mathcal{N}_r^2(u)),
\end{eqnarray}
and
\begin{eqnarray}\label{tahr+}
\mathcal{T}_{AH^+}&=&\frac{1}{2\pi}\left(\frac{M(u)}{(M(u)+\sqrt{M^2(u)-\mathcal{N}_r(u)})^{2}} -\frac{\mathcal{N}_r(u)}{(M(u)+\sqrt{M^2(u)-\mathcal{N}_r(u)})^{3}}\right)\nonumber\\&=& \frac{1}{2\pi}\left(\frac{1}{4M(u)}-{\frac {1}{64}}\,{\frac {{\mathcal{N}^{2}_r}(u)}{{M}^{5}(u)}}\right)+\mathcal{O}(\mathcal{N}_r^3(u)).
\end{eqnarray}
Regarding (\ref{tahr-}) and (\ref{tahr+}), one realizes that the temperature of the inner AH (\ref{lol}) is always negative whereas the temperature of the outer AH (\ref{hol}) is positive. This situation for the temperatures of the  AHs is  exactly similar to what happens for the Reissner-Nordström black hole where for the region   $r_{AH^-}<r<r_{AH^+}$,
the radial coordinate $r$  is timelike and time coordinate $t$ is spacelike. One may write the geodesic equation and see that the gravity force is repulsive at $r=r^{-}_{AH}$ for the Reissner-Nordström black hole. That is, the inner horizon is repelling stuff while outer horizon is attracting them. In fact,
the horizon temperature is related to surface gravity which is the gravitational acceleration of an object falling to the black hole (as measured by the asymptotic observer). Positivity of temperature then means that gravity force is attractive
and vice versa. In our case, this point can be realized by the surface gravity
given by (\ref{surr}).  Here, the negative contribution of the background radiation field, playing the same role  as the charge in the Reissner-Nordström black hole, leads to a negative surface gravity for small $r$ values where
the inner AH (\ref{lol}) lies there. This negative surface gravity yields the gravitational
repulsion and the negative temperature (\ref{tahr-}) for the inner AH (\ref{lol}). One can also discuss
about the gravitational repulsion of the inner AH (\ref{lol}),  preventing anything to reach it,
and its spacelike nature obtained in (\ref{dsahr-}).
Also regarding (\ref{tahempty}) and (\ref{tahr+}),  we see that the temperature of the outer AH (\ref{hol})  of the black hole in the radiation
background is lower than the temperature of the  AH  (\ref{ahempty}) in an empty background. 

The semiclassical Bekenstein-Hawking entropy at the AHs (\ref{lol}) and (\ref{hol}) are
 given
by
\begin{equation}\label{sahr-}
\mathcal{S}_{AH^{-}}=\pi r_{AH^{-}}^2=\pi\left(
\frac{\mathcal{N}_r^2(u)}{4M^{2}(u)} \right)+\mathcal{O}\left(\mathcal{N}_r^3(u)\right),
\end{equation}
and
\begin{equation}\label{sahr+}
\mathcal{S}_{AH^{+}}=\pi r_{AH^{+}}^2=\pi\left( 4M^{2}(u)-2\mathcal{N}_r(u)\right)+\mathcal{O}\left(\mathcal{N}_r^2(u)\right).
\end{equation} 
Comparing (\ref{sahr-}) and (\ref{sahr+}), the inner AH (\ref{lol})  has a negligible entropy relative to the entropy associated with the outer AH (\ref{hol}) in a dilute radiation background. Also regrading (\ref{sahr+}), we see that the background radiation field contributes negatively in the
entropy of the outer AH leading to a lower entropy compared to the entropy
(\ref{sempty}) of the single AH (\ref{ahempty}) in an empty background.
In the limit of $\mathcal{N}_r(u)\rightarrow 0$, the entropy  of the inner AH (\ref{lol}) vanishes as the horizon itself vanishes, while the entropy (\ref{sahr+}) of the outer AH (\ref{hol}) reduces to the entropy (\ref{sempty}) of the
evaporating black hole in an empty space.

\subsection{ Evaporating Vaidya Black Hole Surrounded by the Quintessence Field} 
For the case of Vaidya black hole surrounded by the quintessence field with the equation of state parameter  $\omega_q=-\frac{2}{3}$ \cite{Kiselev, Vik},
the metric (\ref{Vaidya}) takes the following form

\begin{equation}\label{q}
dS^{2} =-\left(1-\frac{2M(u)}{r} -N_{q}(u) r \right)du^2
+2dudr+r^2
d\Omega_{2}^2,
 \end{equation} 
where $N_q(u)$ is the normalization parameter for the quintessence field surrounding the black hole, with
the dimension of $[N_{q}] = l^{-1}$. Regarding the positive energy condition on the surrounding quintessence field,
represented by the relation (\ref{WEC**}), it is required that $N_q(u)>0$. Regarding
the metric (\ref{q}), one realizes the non-trivial contribution of the characteristic feature
of the surrounding quintessence field  to the metric of the Vaidya black hole. In this case, the background quintessence filed
changes the causal structure of this black hole solution in comparison to the original Vaidya black hole in an empty background.
A rather similar effect happens when one immerses a Vaidya black hole in a de Sitter background \cite{ahmallet1} with the difference that here our spacetime
 tends asymptotically to the quintessence rather than  the de Sitter asymptotics.

The surface gravity associated to the metric (\ref{q}) is given by
\begin{equation}\label{surq}
\mathcal{K}=\frac{1}{2}\left(\frac{2M(u)}{r^{2}} -N_q(u) \right).
\end{equation}
Then, we see that the quintessence background appears with a negative surface gravity contribution which decreases the gravitational attraction of the black hole. This characteristic feature of the quintessence  field, namely the gravitational
repulsion, is favored by its cosmological   application as one of the candidates for
the dark energy which is responsible for the current accelerating expansion
of the universe \cite{Vik}. 

The expansion scalar for the outgoing null geodesics reads as
\begin{equation}\label{expq}
\Theta=\frac{1}{r}\left(1-\frac{2M(u)}{r} -N_{q}(u)r\right).
\end{equation}
Then, one can find the  solutions to $\Theta=0$ determining the locations
of the AH
as  
\begin{equation}\label{nini}
r_{1}(u) = \frac{1-\sqrt{1-8 M(u) N_{q}(u)}}{2 N_{q}(u)},
\end{equation}
and
\begin{equation}\label{mini}
r_{2}(u) = \frac{1+\sqrt{1-8 M(u) N_{q}(u)}}{2 N_{q}(u)}.
\end{equation}
Then, depending on the sign of  the  discriminant $\Delta(u)=1-8 M(u) N_{q}(u)$, we have the following
three situations. 
\begin{itemize}
\item For $\Delta(u)<0$, we have no real solutions. Then, there is no physical
AH.

\item For the extremal case of $\Delta(u)=0$, equivalent to $N_{q}(u)=\frac{1}{8M(u)}$,
the solutions (\ref{nini}) and (\ref{mini}) coincide to a single extremal AH, i.e $r_{1}(u) =r_{2}(u)=4M(u)$. 
\item For $\Delta(u)>0$, there are two real physical AHs as $r_{AH^{-}}(u)=r_1
(u)$
and $r_{AH^{+}}(u)=r_2(u)$ for the black hole in the quintessence background in contrast to the black hole in the empty
space and in the dust field background having a single AH. 
\end{itemize}

For the extremal case of $\Delta(u)=0$, there is only one single  extremal
AH given by
 \begin{equation}\label{ext}
r_{AH^{ext}}(u)= r_{1}(u) =r_{2}(u)=4M(u), 
\end{equation} which is twice of the dynamical Schwarzschild radius. Then, the dynamics of the single extremal AH (\ref{ext}) is governed by the relation 
\begin{equation}\label{ah*q}
\dot r_{AH^{ext}}(u)=-4 L.
\end{equation}
This shows that the single extremal AH (\ref{ext}) for the black hole in the quintessence
background  is moving inward during the evaporation, i.e
it is shrinking.
 For this case, the nature of the extremal AH (\ref{ext}) can be determined by
looking at the induced metric on it as
\begin{equation}\label{0}
dS_{{AH}^{ext}}^2 =-8L\,du^2+16M^2(u)
d\Omega_{2}^2,
 \end{equation}
showing that the extremal AH (\ref{ext})  is a timelike surface for an evaporating BH in the quintessence background.   
 The Hawking temperature on the extremal AH (\ref{ext}) can be determined using the surface gravity as
\begin{equation}\label{tahextq*}
\mathcal{T}_{AH^{ext}}=\frac{1}{4\pi}\left(\frac{2M(u)}{ r_{AH^{ext}}^2}-\frac{1}{8M(u)}\right)=0,
\end{equation}
indicating that the extremal AH (\ref{ext}) is a zero temperature surface.
 One may interpret this zero temperature as the suppressed or zero Hawking radiation for the extremal black hole in the quintessence background.
 Finally, the semiclassical Bekenstein-Hawking entropy at AH (\ref{ext}) is given
by
\begin{equation}
\mathcal{S}_{AH^{ext}}=\pi r_{AH^{ext}}^2=16\pi M^2(u).
\end{equation}
Then, the entropy of the extremal AH (\ref{ext}) is four times the entropy
(\ref{sempty}) associated to the single AH (\ref{ahempty}) of the black hole in an empty space.

The location of two AHs (\ref{nini}) and (\ref{mini})  for the case $\Delta(u)>0$ with a small  quintessence parameter, i.e $N_q(u)\ll
M(u)$,
are given by
\begin{equation}\label{gogo}
r_{AH^-}(u)=
2M(u)+4M^{2}(u)N_{q}(u) +\mathcal{O}\left(N_{q}^2(u)\right),
\end{equation}
and
\begin{equation}\label{mago}
r_{AH^+}(u)=
\frac{1}{N_{q}(u)}-2M(u)-4{M}^{2}(u)N_{q}(u)+\mathcal{O}\left(N_{q}^2(u)\right).
\end{equation}
 It is seen that the contribution of the surrounding background quintessence field leads
to ($i$) the existence of an inner AH (\ref{gogo}) larger than the dynamical Schwarzschild radius $r(u)=2M(u)$, and ($ii$) a large outer (cosmological) AH  (\ref{mago})  tending to infinity for $N_q(u)\lll1$. Regarding (\ref{gogo}) and (\ref{mago}), the quintessence field has a positive contribution to the inner AH while
the black hole mass has a negative contribution to the outer AH.  Then, as the black hole evaporation continues, the inner AH   (\ref{gogo}) shrinks to vanish while the outer AH  (\ref{mago})  expands toward its asymptotic value of $ N^{-1}_{q}(u)$.
  
For this case, the dynamics of the AHs (\ref{gogo})
and (\ref{mago}) are given by 
\begin{equation}\label{jiji}
\dot r_{AH^{-}}(u) =-2L-8LM(u)N_{q}(u) +4M^{2}(u)\dot N_{q}(u)+\mathcal{O}\left(N_{q}^2(u)\right) ,
\end{equation}
and 
\begin{equation}\label{giji}
\dot r_{AH^{+}}(u) = 2L+8LM(u)N_{q} -\frac{\dot N_q (u)}{N^2_{q}(u)}-4{M}^{2}(u)\dot N_{q}(u)+\mathcal{O}\left(N_{q}^2(u)\right),
\end{equation}
respectively. 
Then one realizes the following points:
\begin{itemize}
\item For $\dot N_{q}(u)=0$ representing the static surrounding quintessence field,  the inner and outer AHs (\ref{gogo})
and (\ref{mago}) have equal velocities but with opposite signs. Then, the inner AH (\ref{gogo}) shrinks while the outer (\ref{mago})  expands
during the evaporation. Also, we observe that both of these
AHs are moving faster than the case of single AH (\ref{ahempty}) for the black hole in an empty space.

 \item For $\dot N_{q}(u)<0$, the inner AH (\ref{gogo})  shrinks while the outer (\ref{mago}) expands, and these are faster than the cases in static quintessence background and in empty spaces. 
\item For $\dot N_{q}(u)>0$, the behaviour of the AHs (\ref{gogo}) and (\ref{mago}) depend on the $M(u)$,
$L$, $N_q(u)$ and $\dot N_q(u)$ values.
For the specific case of $4M^{2}(u)\dot N_{q}(u)=2L+8LM(u)N_{q}(u)$, the
inner AH (\ref{gogo}) is frozen while the outer AH (\ref{mago}) is moving inward.
\end{itemize}
 The nature of the AHs (\ref{gogo}) and (\ref{mago}) can be determined by
looking at the induced metrics on them up to $\mathcal{O}(N_q)$ as
\begin{equation}\label{natureq-}
{dS^{2}_{AH^-}}=2\left( -2L-8LM(u)N_{q}(u) +4M^{2}(u)\dot N_{q}(u)\right)du^2+r_{AH^-}^{2}
d\Omega_{2}^2,
 \end{equation}
and 
\begin{equation}\label{natureq+}
{dS^{2}_{AH^+}} =2\left( 2L+8LM(u)N_{q}-\frac{\dot N_q (u)}{N^2_{q}(u)} -4{M}^{2}(u)\dot N_{q}(u)\right) du^2+r_{AH^+}^{2}
d\Omega_{2}^2.
 \end{equation}
These show that for $\dot N_{q}(u)\leq0$,  the inner AH (\ref{gogo})  is a timelike surface while the outer AH (\ref{gogo})
is an spacelike surface. For $\dot N_{q}(u)>0$, the nature  of these AHs depend on the $M(u)$,
$L$, $N_q(u)$ and $\dot N_q(u)$ values. For the specific case of $4M^{2}(u)\dot N_{q}(u)=2L+8LM(u)N_{q}(u)$, the
inner AH (\ref{gogo}) is null while the outer (\ref{mago}) is timelike.

The Hawking temperature on the AHs (\ref{gogo}) and (\ref{mago}) of the black hole are
\begin{equation}
\mathcal{T}_{AH^{-}}=\frac{1}{4\pi}\left(\frac{8M(u)N^2_{q}(u)}{\left(1-\sqrt{1-8  M(u)N_{q}(u)}\right)^{2}} -N_q(u) \right)= \frac{1}{4\pi}\left(\frac{1}{2M(u)}-3N_{q}(u) \right)
+\mathcal{O}\left(N^2_{q}(u)\right),
\end{equation}
and 
\begin{equation}
\mathcal{T}_{AH^{+}}=\frac{1}{4\pi}\left(\frac{8M(u)N^2_{q}(u)}{\left(1+\sqrt{1-8  M(u)N_{q}(u)}\right)^{2}} -N_q(u) \right)= \frac{1}{4\pi}\left(-N_{q}(u)+2M(u)N^2_{q}(u) \right)
+\mathcal{O}\left(N^3_{q}(u)\right).
\end{equation}
Then, regarding the condition $8 M(u) N_{q}(u)<1$, one realizes that the temperature of the inner AH  (\ref{gogo}) is always positive while
the temperature of the outer AH (\ref{mago})  is negative. For the inner AH (\ref{gogo}), the background quintessence field decreases
the  temperature of the  AH relative to the case of black hole in empty
space. Here, regarding (\ref{surq}), the negative temperature of the outer AH (\ref{mago}) is resulting from the
 fact that the quintessence background field effect dominates at large distances
which leads to the negative
surface gravity, namely the gravitational repulsion. 

 Finally, the semiclassical Bekenstein-Hawking entropy at the horizons (\ref{gogo}) and (\ref{mago}) are given by
\begin{equation}\label{sq-}
\mathcal{S}_{AH^{-}}=\pi r_{AH^{-}}^{2}=\pi\left(\frac{1-\sqrt{1-8 N_{q}(u) M(u)}}{2 N_{q}(u)}\right)^2= \pi \left(4M^2(u) +16M^3(u) N_{q}(u)\right)+\mathcal{O}\left(N^2_{q}(u)\right),
\end{equation}
and 
\begin{equation}\label{sq+}
\mathcal{S}_{AH^{+}}=\pi r_{AH^{+}}^{2}=\pi\left(\frac{1+\sqrt{1-8 N_{q}(u) M(u)}}{2 N_{q}(u)}\right)^2=\pi\left(\frac{1}{N_q ^2(u)}-\frac{4M(u)}{N_q(u)}-4M^{2}(u)-16M^{3}(u)N_{q}(u)\right)
+\mathcal{O}\left(N^2_{q}(u)\right).
\end{equation}
Then, for the quintessence background with $N_q(u)\ll1$,  the inner AH (\ref{gogo}) has a negligible entropy in comparison  to the entropy (\ref{mago}) associated with the outer AH. Also regrading (\ref{sq-}), we see that the background quintessence field contributes positively in the
entropy of the inner AH (\ref{gogo}), which leads  to a higher entropy relative to the single AH (\ref{ahempty}) of
the black hole in an empty background.
In the limit of $N_q(u)\rightarrow 0$, the entropy (\ref{sq-}) of the inner AH (\ref{gogo}) tends to the entropy (\ref{sempty}) of AH (\ref{ahempty}) of
the black hole in an empty background while the entropy of the outer AH blows
up to infinity, as the horizon itself goes to infinity.
Also regarding (\ref{sq+}), one realizes that since the entropy is a non-negative quantity, then it is required that the first term overcomes to all of the next negative terms. This requires  the
value of quintessence normalizing parameter $N_q(u)$ to be very small.
In fact,
this entropy consideration on the cosmological AH (\ref{mago}) gives a direct restriction
on the quintessence field normalization parameter $N_q(u)$.

\subsection{ Evaporating Vaidya Black Hole Surrounded by the Cosmological Constant-Like  Field } 
 For the case of Vaidya black hole surrounded by the cosmological constant-like  field with the equation of state parameter
$\omega_c = -1$ \cite{Kiselev, Vik},  which we call it in the following as the cosmological field in short, the metric (\ref{Vaidya}) takes the following form
\begin{equation}\label{0**}
dS^{2} =-\left(1-\frac{2M(u)}{r} -N_{c}(u) r^2 \right)du^2
+2dudr+r^2
d\Omega_{2}^2,
 \end{equation} 
where $N_c(u)$ is the normalization parameter for the cosmological field surrounding the black hole, with
the dimension of $[N_c(u)] = l^{-2}$. Regarding the positive energy condition on the surrounding cosmological   field,
represented by the relation (\ref{WEC**}), it is required that $N_c(u)>0$. Regarding
the metric (\ref{0**}), one realizes the non-trivial contribution of the characteristic feature
of the surrounding cosmological  field  to the metric of the Vaidya black hole. In this case, the background cosmological field
changes the causal structure of the original Vaidya black hole in an empty background to the Vaidya-de Sitter space \cite{ahmallet1}, similar to the case of Schwarzschild to Schwarzschild-de Sitter case. Notice to the difference
between our more general solution here and that of obtained in \cite{ahmallet1} having a constant cosmological field.

 The  surface gravity associated with the metric (\ref{0**}) is given by 
\begin{equation}\label{surc}
\mathcal{K}=\frac{M(u)}{r^{2}} -N_c(u) r .
\end{equation}
Then, we see that the cosmological  background appears with a negative surface gravity contribution  which decreases the gravitational attraction of the black hole. This characteristic feature of the cosmological
field, namely the gravitational repulsion, is evidenced by its application as the
most favored candidates for the
dark energy as the field responsible for the current accelerating expansion of the universe \cite{Vik}.

The expansion scalar for the outgoing null geodesics reads as
\begin{equation}\label{expc}
\Theta=\frac{1}{r}\left(1-\frac{2M(u)}{r} -N_{c}(u)\,r^2\right).
\end{equation}
Then, one can find the  solutions to  $\Theta=0$ determining the locations
of the AHs
as  
\begin{eqnarray}
&&r_{1}(u)=-\frac{1}{3^{\frac{1}{3}}  \left(9M(u)N_c^2(u)+ \sqrt{-3N_c^3(u) \Delta(u)}  \right)^\frac{1}{3}}
-\frac{\left(9M(u)N_c^2(u)+\sqrt{-3N_c^3(u) \Delta(u)}   \right)^\frac{1}{3}}{3^\frac{2}{3}N_c(u)},\nonumber\\
&&r_2(u)=\frac{1-i\sqrt 3}{2\times 3^{\frac{1}{3}}  \left(9M(u)N_c^2(u)+ \sqrt{-3N_c^3(u) \Delta(u)}  \right)^\frac{1}{3}}
+\frac{\left(1+i\sqrt 3\right)\left(9M(u)N_c^2(u)+ \sqrt{-3N_c^3(u) \Delta(u)}   \right)^\frac{1}{3}}{2\times 3^\frac{2}{3}N_c(u)},\nonumber\\
&&r_{3}(u)=\frac{1+i\sqrt 3}{2\times3^{\frac{1}{3}}  \left(9M(u)N_c^2(u)+ \sqrt{-3N_c^3(u) \Delta(u)}  \right)^\frac{1}{3}}
+\frac{\left(1-i\sqrt 3\right)\left(9M(u)N_c^2(u)+ \sqrt{-3N_c^3(u) \Delta(u)}   \right)^\frac{1}{3}}{2\times 3^\frac{2}{3}N_c(u)},
\end{eqnarray}
where $\Delta(u)=1-27M^{2}(u)N_c(u)$. Then, depending on the sign of  the  discriminant $\Delta(u)$ we have the following
three situations:
\begin{itemize}
\item For $\Delta(u)<0$, the first solution is negative 
while the second and third solutions are imaginary and complex conjugate.
Then, there is no physical AH.
\item For the extremal case $\Delta(u)=0$, equivalent to $N_c(u)=\frac{1}{27M^2(u)}$,
the first solution will be negative as $r_1=-\frac{2}{\sqrt{3N_c(u)}}$ and the second and third solutions coincide to a single extremal AH as $r_2(u)=r_3(u)=\frac{1}{\sqrt{3N_c(u)}}=3M(u)$.
Then, the first solution is not a physical horizon while the coinciding second
and third solutions  are representing a physical
extremal AH. 
\item For $\Delta(u)>0$,  the first solution will
be  negative, and gives no physical AH  while the second
and third solutions give two physical inner and outer AHs.
\end{itemize}
For the extremal case of $\Delta(u)=0$, there is only one single  extremal
AH given by \begin{equation}\label{extc}
r_{AH^{ext}}(u)= r_{AH^{-}}(u) =r_{AH^{+}}(u)=3M(u), 
\end{equation} which is one and a half times of the dynamical Schwarzschild radius. Then, the dynamics of the single extremal AH  (\ref{extc}) is governed by the relation 
\begin{equation}\label{ah*c}
\dot r_{AH^{ext}}(u)=-3 L.
\end{equation}
This shows that the single extremal AH  (\ref{extc}) for the black hole in the cosmological
background  is moving inward during the evaporation, i.e
it is shrinking.
 For this case, the nature of the extremal AH (\ref{extc}) can be determined by
looking at the induced metric on it as
\begin{equation}\label{0c}
dS_{{AH}^{ext}}^2 =-6L\,du^2+16M^2(u)
d\Omega_{2}^2,
 \end{equation}
indicating that the extremal AH  (\ref{extc})  is a timelike surface for an evaporating BH in the cosmological background.   
 The Hawking temperature on the extremal AH (\ref{extc})   can be determined using the surface gravity as
\begin{equation}\label{tahextc}
\mathcal{T}_{AH^{ext}}=\frac{1}{2\pi}\left(\frac{M(u)}{ r_{AH^{ext}}^2}-3M(u)r_{AH^{ext}}\right)=0,
\end{equation}
indicating that the extremal AH  (\ref{extc})  is a zero temperature surface.
 One may interpret this zero temperature as the suppressed or  zero Hawking radiation for
 the extremal black hole in the cosmological background.
 Finally, the semiclassical Bekenstein-Hawking entropy at extremal AH (\ref{extc}) is given
by
\begin{equation}
\mathcal{S}_{AH^{ext}}=\pi r_{AH^{ext}}^2=9\pi M^2(u).
\end{equation}
Then, the entropy of the extremal AH  (\ref{extc})  is $\frac{9}{4}$ the entropy (\ref{sempty}) associated to the single AH (\ref{ahempty}) of the black hole in an empty space.

For the case of $\Delta(u)>0$, we have the following solutions
\begin{equation}\label{lili}
r_{AH^{-}}(u)=r_2 (u) =
2M(u)+8M^{3}(u)N_{c}(u) +\mathcal{O}\left(N_{c}^{2}(u)\right),
\end{equation}
\begin{equation}\label{lala}
r_{AH^{+}}(u) =r_3(u)=
\frac{1}{\sqrt{N_{c}(u)}}-M(u)-\frac{3}{2}{M}^{2}(u)\sqrt{N_{c}(u)}-4M^3(u) N_c(u)
+\mathcal{O}\left(N_{c}^{\frac{3}{2}}(u)\right),
\end{equation}
representing the  inner and outer  AH, respectively. It is seen that the contribution of the surrounding background cosmological field leads
to ($i$) the existence of an inner AH (\ref{lili}) larger than the dynamical Schwarzschild radius
$r(u)=2M(u)$, and ($ii$) a large outer (cosmological) AH (\ref{lala}) tending to infinity for $N_c(u)\lll1$. Regarding (\ref{lili}) and (\ref{lala}), similar to the quintessence case, the cosmological
field has a positive contribution to the inner AH while the black hole mass has a negative contribution to
the outer AH.  Then, as the black hole evaporation continues, the inner AH
(\ref{lili}) shrinks to
vanish while the outer AH (\ref{lala}) expands toward its asymptotic value of $ N^{-\frac{1}{2}}_{c}(u)$.

   For this case, the dynamics of the AHs (\ref{lili})
and (\ref{lala}) are given by
\begin{equation}
\dot r_{AH^{-}}(u)=
-2L-24L M^{2}(u)N_{c}(u)+8M^{3}(u)\dot N_{c}(u) +\mathcal{O}(N_{c}^2), 
\end{equation}
and 
\begin{equation}\label{rah+c}
\dot r_{AH^{+}}(u)=
L+3L M(u)\sqrt N_{c}+12L M^2(u) N_c-\frac{\dot N_c(u)}{2N_c^{\frac{3}{2}}}-\frac{3}{4}{M}^{2}(u)\frac{\dot N_{c}(u)}{\sqrt{N_c(u)}}-4M^3(u) \dot N_c(u)
+\mathcal{O}(N_{c}^{\frac{3}{2}}),
\end{equation}
respectively. Then, one realizes the following points:
\begin{itemize}
\item For $\dot N_{c}(u)=0$, showing the static surrounding cosmological field, the inner AH (\ref{lili}) shrinks while the outer AH (\ref{lala}) expands. Also, we observe
that the inner AH (\ref{lili}) is moving inward faster than the case of AH
(\ref{ahempty}) for the black hole in an empty space. \item For $\dot N_{c}(u)<0$, the inner AH (\ref{lili}) shrinks while the outer AH (\ref{lala})  expands,
and these are faster than the cases in static cosmological field background and
in empty space. 
\item For $\dot N_{c}(u)>0$, the behaviour of these AHs (\ref{lili}) and
(\ref{lala}) depend on the $M(u)$,
$L$, $N_c(u)$ and $\dot N_c(u)$ values.
For the specific case of $8M^{3}(u)\dot N_{c}(u)=2L+24L M^{2}(u)N_{c}(u)$,
the inner AH (\ref{lili})  is frozen while the outer AH (\ref{lala}) dynamics depends on the total value of the remaining terms in (\ref{rah+c}).
\end{itemize}

 The
nature of  AHs (\ref{lili}) and (\ref{lala}) can be determined by looking at the induced metrics on them up to $\mathcal{O}(N_{c})$ as
 \begin{equation}\label{nature1}
{dS^{2}_{AH^-}}=2
\left( -2L-24L M^{2}(u)N_{c} +8M^{3}(u)\dot N_{c}(u) \right)du^2+r_{AH^-}^{2}
d\Omega_{2}^2,
 \end{equation}
and 
\begin{equation}\label{nature2}
{dS^{2}_{AH^+}}=2\left( L+3L M(u)\sqrt N_{c}+12L M^2(u) N_c-\frac{\dot N_c(u)}{2N_c^{\frac{3}{2}}}-\frac{3}{4}{M}^{2}(u)\frac{\dot N_{c}(u)}{\sqrt{N_c(u)}}-4M^3(u) \dot N_c(u)
 \right)du^2+r_{AH^+}^{2}
d\Omega_{2}^2.
 \end{equation}
These show that for $\dot N_{c}(u)\leq0$,  the inner AH (\ref{lili}) is a timelike surface while the outer AH (\ref{lala})
is spacelike surface. For $\dot N_{c}(u)>0$, the nature  of these AHs depends on the $M(u)$, $L$, $N_c(u)$ and $\dot N_c(u)$ values.
For the specific case of $8M^{3}(u)\dot N_{c}(u)=2L+24L M^{2}(u)N_{c}(u)$,
the inner AH (\ref{lili}) is null while the nature of the outer AH (\ref{lala})
depends on the total value of the remaining terms in the parenthesis in (\ref{nature2}).

The Hawking temperature on the inner and outer AHs (\ref{lili}) and  (\ref{lala})
of black hole can be determined using the surface gravity as
\begin{eqnarray}
\mathcal{T}_{AH^{-}}&=& \frac{1}{2\pi}\left(\frac{M(u)}{\left(2M(u)+8M^3(u)N_c(u) +\mathcal{O}\left(N_c^2(u)\right)\right)^2} -N_c(u)\left(  2M(u)+8M^2(u)N_c(u) +\mathcal{O}\left(N_c^2(u)\right)\right)  \right)\nonumber\\
&=&\frac{1}{2\pi}\left(\frac{1}{4}\frac{1}{M(u)}-4\,M(u)N_c(u)
\right)+\mathcal{O}\left(N^2_{c}(u)\right),
\end{eqnarray}
and
\begin{eqnarray}
\mathcal{T}_{AH^{+}}&=& \frac{1}{2\pi}\left(\frac{M(u)}{\left( \frac{1}{\sqrt{ N_{c}(u)}}-M(u)-\frac{3}{2}{M}^{2}(u)\sqrt{N_{c}(u)}-4M^3(u) N_c(u)
+\mathcal{O}\left(N_{c}^{\frac{3}{2}}(u)\right)\right)^2}\right)\nonumber\\ &&-\frac{1}{2\pi}\left(N_c(u)\left(  \frac{1}{\sqrt{N_{c}(u)}}-M(u)-\frac{3}{2}{M}^{2}(u)\sqrt{N_{c}(u)}+\mathcal{O}(N_c(u))\right)  \right)\nonumber\\
&=& \frac{1}{2\pi}\left(-\sqrt{{N_c}(u)}+2\,M(u)N_c(u)
\right)+\mathcal{O}\left(N^\frac{3}{2}_{c}(u)\right),
\end{eqnarray}
respectively.
One realizes that the temperature of the inner AH (\ref{lili})  is always
positive while the temperature of the outer AH (\ref{lala})  is negative.
For the inner AH (\ref{lili}), the background cosmological 
field decreases the temperature of the AH relative to the AH (\ref{ahempty})   of black hole in empty space. Here, regarding
(\ref{surc}), the negative temperature of the outer AH  (\ref{lala})  is resulting from the fact that the cosmological background
field effect dominates at large distances which leads to the negative surface gravity, equivalently to the gravitational repulsion.

Finally, the semiclassical Bekenstein-Hawking entropy at the AHs (\ref{lili})
and (\ref{lala}) are given
by
\begin{equation}\label{sahc-}
\mathcal{S}_{AH^{-}}=\pi r_{AH^{-}}^{2}= \pi \left(4M^2(u) +32M^4(u) N_{c}(u)\right)+\mathcal{O}\left(N^2_{c}(u)\right),
\end{equation}
and 
\begin{eqnarray}\label{sahc+}
\mathcal{S}_{AH^{+}}&=&\pi r_{AH^{+}}^{2}\nonumber\\
&=&\pi\left(\frac{1}{N_c(u)}-2\,{\frac{M(u)}{\sqrt{N_c(u)}}}-2\,{M}^{2}(u)-5\,{M}^{3}(u)
\sqrt{N_c(u)}-16\,{M}^{4}(u)N_c(u)
\right)+\mathcal{O}\left(N_{c}^{\frac{3}{2}}(u)\right).
\end{eqnarray}
Then, for the cosmological field background with $N_c(u)\lll 1$,  the inner AH (\ref{lili})   has a negligible entropy (\ref{sahc-}) in comparison  to the entropy
(\ref{sahc+})
 associated with the outer AH (\ref{lala}). Also regrading (\ref{sahc-}), we see that the background cosmological field contributes positively to the
entropy of the inner AH (\ref{lili}), which leads  to a higher entropy relative to the single AH (\ref{ahempty}) of
the black hole in an empty background.
In the limit of $N_c(u)\rightarrow 0$, the entropy (\ref{sahc-}) of the inner AH (\ref{lili}) tends to the entropy (\ref{sempty}) of AH (\ref{ahempty})
of the black hole in an empty background while the entropy (\ref{sahc-})  of the outer AH (\ref{lala}) blows up to infinity.
Also regarding (\ref{sahc+}), one realizes that since the entropy is a non-negative quantity, then this relation requires that the first term overcomes to all of the next negative terms. This requires the
value of cosmological field normalizing parameter $N_c(u)$ to be very small.
In fact, this entropy consideration on the cosmological AH (\ref{lala}) gives a restriction on the cosmological field normalization parameter $N_c(u)$.


\subsection{ Evaporating Vaidya Black Hole Surrounded by the Phantom Field} 
For the case of Vaidya black hole surrounded by the phantom field with the equation of state parameter
$\omega_p = -\frac{4}{3}$ \cite{Kiselev, Vik}, the metric (\ref{Vaidya}) takes the following form
\begin{equation}\label{-2/3}
dS^{2} =-\left(1-\frac{2M(u)}{r} -N_p(u) r^{3} \right)du^2
+2dudr+r^2
d\Omega_{2}^2,
 \end{equation} 
where $N_p(u)$ is the normalization parameter for the phantom field surrounding the black hole, with
the dimension of $[N_{p}(u)] = l^{-3}$. Regarding the positive energy condition on the surrounding phantom  field,
represented by the relation (\ref{WEC**}), it is required that $N_p(u)>0$. Regarding
the metric (\ref{-2/3}), one realizes the non-trivial contribution of the characteristic feature
of the surrounding phantom field  to the metric of the Vaidya black hole. In this case, the background phantom filed
changes the causal structure of the original Vaidya black hole in an empty background to the case with a phantom asymptotic.

The  surface gravity associated with the metric (\ref{-2/3}) is given by  
\begin{equation}\label{surp}
\mathcal{K}=\frac{1}{2}\left(\frac{2M(u)}{r^{2}} -3N_p(u) r^2\right) .
\end{equation}
Then, we see that the phantom background appears with a negative surface gravity contribution which decreases the gravitational attraction of the black hole. This characteristic feature of the phantom  field, namely the gravitational
repulsion, is favored by its cosmological   application as one of the candidates for
the dark energy \cite{Vik}. 

The expansion scalar for the outgoing null geodesics reads as
\begin{equation}\label{expp}
\Theta=\frac{1}{r}\left(1-\frac{2M(u)}{r} -N_{p}(u)\,r^3\right).
\end{equation}
Then, we find the  solutions of  $\Theta=0$ determining the locations
of the AH
as  
\begin{eqnarray}
&&r_{1}(u)=-\frac{1}{2}\sqrt{\Gamma(u)}-\frac{1}{2}\sqrt{-\Gamma(u)-\frac{2}{N_{p}(u)\sqrt{\Gamma(u)}}},\nonumber\\
&&r_2(u)=-\frac{1}{2}\sqrt{\Gamma(u)}+\frac{1}{2}\sqrt{-\Gamma(u)-\frac{2}{N_{p}(u)\sqrt{\Gamma(u)}}},\nonumber\\
&&r_{3}(u)=\frac{1}{2}\sqrt{\Gamma(u)}-\frac{1}{2}\sqrt{-\Gamma(u)+\frac{2}{N_{p}(u)\sqrt{\Gamma(u)}}},\nonumber\\
&&r_{4}(u)=\frac{1}{2}\sqrt{\Gamma(u)}+\frac{1}{2}\sqrt{-\Gamma(u)+\frac{2}{N_{p}(u)\sqrt{\Gamma(u)}}},
\end{eqnarray}
where 
\begin{equation}
\Gamma(u)=\frac{8\left( \frac{2}{3}\right)^{\frac{1}{3}}M(u)}{\left(9N_p(u)
-\sqrt{81N_p^2(u)\Delta(u)}  \right)^{\frac{1}{3}}}+\frac{\left(9N_p(u)
-\sqrt{81N_p^2(u)\Delta(u)}  \right)^{\frac{1}{3}}}{2^\frac{1}{3} 3^\frac{2}{3}
N_p(u)},
\end{equation}
and $\Delta(u)=1-\frac{2048}{27}M^3(u)N_p(u)$.
Then, depending on the sign of  the  discriminant $\Delta(u)$ we have the following three situations. 
\begin{itemize}
\item For $\Delta(u)<0$, the first and second solutions are negative 
while the third and fourth solutions are imaginary.
Then, there is no physical AH.
\item For the extremal case $\Delta(u)=0$, equivalent to $N_p(u)=\frac{27}{2048M^3(u)}$, the first and second solutions coincide as $r_1(u)=r_{2}(u)=\frac{8}{3}\left(\frac{27}{2048
N_p(u)}
\right)^{\frac{1}{3}}
$ while the third and fourth solutions read as  $r_3(u)=-r_4(u)=\frac{8}{3} i \left(\sqrt{2}+i\right) M(u)$.
Then, the first and second  coinciding solutions are physical AH while the third and fourth coinciding solutions  are not physical.
\item For $\Delta(u)>0$,  the first and second solutions are imaginary and
complex conjugate to each other and gives no physical AH,  while the third
and fourth solutions give two physical inner and outer AH. 
\end{itemize}
For the extremal case of $\Delta(u)=0$, there is only one single  extremal
AH given by \begin{equation}\label{extp}
r_{AH^{ext}}(u)=r_1(u)=r_{2}(u)=\frac{8}{3}M(u), 
\end{equation} which is $4/3$  of the dynamical Schwarzschild radius. Then, the dynamics of the single extremal AH (\ref{extp}) is governed by the relation 
\begin{equation}\label{ah*p}
\dot r_{AH^{ext}}(u)=-\frac{8}{3} L.
\end{equation}
This shows that the single extremal AH (\ref{extp})   for the black hole in the quintessence
background  is moving inward during the evaporation, i.e
it is shrinking.
 For this case, the nature of the extremal AH (\ref{extp}) can be determined by
looking at the induced metric on it as
\begin{equation}\label{0p}
dS_{{AH}^{ext}}^2 =-\frac{16}{3}L\,du^2+16M^2(u)
d\Omega_{2}^2,
 \end{equation}
indicating that the extremal AH (\ref{extp})  is a timelike surface for an evaporating BH in the phantom background.   
 The Hawking temperature on the extremal AH (\ref{extp})  can be determined using the surface gravity as
\begin{equation}\label{tahextp}
\mathcal{T}_{AH^{ext}}=\frac{1}{4\pi}\left(\frac{2M(u)}{ r_{AH^{ext}}^2}-3\left(\frac{27}{2048M^3(u)}\right)r^{2}_{AH^{ext}}\right)=0,
\end{equation}
indicating that the extremal AH (\ref{extp})  is a zero temperature surface.
 One may interpret this zero temperature as the suppressed or  zero Hawking radiation for
 the extremal black hole in the phantom background.
The semiclassical Bekenstein-Hawking entropy at AH (\ref{extp}) is given
by
\begin{equation}
\mathcal{S}_{AH^{ext}}=\pi r_{AH^{ext}}^2=\frac{64\pi}{9} M^2(u).
\end{equation}
Then, the entropy of the extremal AH (\ref{extp})   is $\frac{16}{9}$ the entropy (\ref{sempty}) associated to the single AH (\ref{ahempty}) of the black hole in an empty space.

For the case of $\Delta(u) > 0$, we have the following solutions
\begin{equation}\label{chichi}
r_{AH^{-}}(u)=r_3 (u) =
2M(u)+16M^{4}(u)N_{p}(u) +\mathcal{O}(N_{p}^2),
\end{equation}
\begin{eqnarray}\label{michi}
r_{AH^{+}}(u) =r_4(u)=
\frac{1}{ N^\frac{1}{3}_{p}(u)}-\frac{2}{3}M(u)-\frac{8}{9}{M}^{2}(u) N^{\frac{1}{3}}_{p}(u)
-\frac{160}{81} M^3(u) N^{\frac{2}{3}}_p(u)-\frac{16}{3}M^4(u) N_p(u)+\mathcal{O}\left(N_{p}^\frac{4}{3}(u)\right),
\end{eqnarray}
representing the  inner and outer  AHs, respectively. It is seen that the contribution of the  background phantom field leads
to ($i$) the existence of an inner AH (\ref{chichi}) larger than the dynamical Schwarzschild radius
$r(u)=2M(u)$, and ($ii$) a large outer (cosmological) AH (\ref{michi}) tending to infinity for $N_p\lll1$. Regarding (\ref{chichi}) and (\ref{michi}), similar to the quintessence and cosmological field cases,  the phantom field has
a positive contribution to the inner AH (\ref{chichi}) while the black hole mass has a negative contribution to the outer
AH. Then, as the black hole evaporation continues, the inner AH (\ref{chichi}) shrinks to vanish while the outer AH (\ref{michi}) expands toward its asymptotic value of $ N^{-\frac{1}{3}}_{p}(u)$.
   For this case, the dynamics of the AHs (\ref{chichi})
and (\ref{michi}) are given by
\begin{equation}
\dot r_{AH^{-}}(u)=
-2L-64L M^{2}(u)N_{p}(u) +16M^{4}(u)\dot N_{p}(u)+ \mathcal{O}\left(N_{p}^2(u)\right), \end{equation}
and 
\begin{eqnarray}\label{dotrp}
\dot r_{AH^{+}}(u)&=&
\frac{2}{3}L+\frac{16}{9}L M(u)N_{p}^{\frac{1}{3}}(u)+\frac{160}{27}L M^2(u) N_p^{\frac{2}{3}}(u)+\frac{64}{3}L M^3(u) N_p(u)\nonumber\\
&&- 
\frac{\dot N_p(u)}{ 3N^\frac{4}{3}_{p}(u)}
-\frac{8}{27}\frac{\dot N_p(u)}{ N^\frac{2}{3}_{p}(u)}{M}^{2}(u)-\frac{320}{243} \frac{\dot N_p(u)}{ N^\frac{1}{3}_{p}(u)}M^3(u)-\frac{16}{3}M^4(u) \dot N_p(u)+\mathcal{O}\left(N_{p}^{\frac{4}{3}}(u)\right),
\end{eqnarray}
respectively. Then one realizes the following points:\begin{itemize}
\item For $\dot N_{p}(u)=0$ representing the static surrounding phantom field, the inner AH (\ref{chichi}) shrinks while the outer AH  (\ref{michi}) expands. Also, we observe
that the inner AH (\ref{chichi})  is moving inward faster than the case of AH (\ref{ahempty}) for the black hole in an empty space. 
\item For $\dot N_{p}(u)<0$, the inner AH (\ref{chichi})  shrinks while the outer AH (\ref{chichi}) expands, and these are faster than the cases in static phantom field background and in empty space. 
\item For $\dot N_{p}(u)>0$, the behaviour of the AHs (\ref{chichi}) and  (\ref{michi}) depend on the $M(u)$,
$L$, $N_p(u)$ and $\dot N_p(u)$ values. For the specific case of $16M^{4}(u)\dot N_{p}(u)=2L+64L M^{2}(u)N_{p}(u) $, the inner AH (\ref{chichi}) is frozen while the outer's
behaviour
depends on the total value of the remaining terms in (\ref{dotrp}).

\end{itemize}
 The nature of the AHs (\ref{chichi}) and (\ref{michi}) can be determined by looking at the induced metric on them up to $\mathcal{O}\left(N_{p}(u)\right)$ as
 \begin{equation}\label{naturep-}
{dS^{2}_{AH^-}}=2
\left( -2L-64L M^{2}(u)N_{p}(u) +16M^{4}(u)\dot N_{p}(u) \right)du^2+r_{AH^-}^{2}
d\Omega_{2}^2,
 \end{equation}
and 
\begin{eqnarray}\label{naturep+}
{dS^{2}_{AH^+}}&=&2(\frac{2}{3}L+\frac{16}{9}L M(u)N_{p}^{\frac{1}{3}}(u)+\frac{160}{27}L M^2(u) N_p^{\frac{2}{3}}(u)+\frac{64}{3}L M^3(u) N_p(u)\nonumber\\
&&-\frac{\dot N_p(u)}{ 3N^\frac{4}{3}_{p}(u)}
-\frac{8}{27}\frac{\dot N_p(u)}{ N^\frac{2}{3}_{p}(u)}{M}^{2}(u)-\frac{320}{243} \frac{\dot N_p(u)}{ N^\frac{1}{3}_{p}(u)}M^3(u)-\frac{16}{3}M^4(u) \dot N_p(u))du^2+r_{AH^+}^{2}
d\Omega_{2}^2.
 \end{eqnarray}
These indicate that for $\dot N_{p}(u)\leq0$,  the inner AH (\ref{chichi})  is a timelike surface while the outer AH (\ref{michi})
is spacelike surface. For $\dot N_{p}(u)>0$, the nature  of these AH depend on the $M(u)$, $L$, $N_p(u)$ and $\dot N_p(u)$ values.
For the specific case of $16M^{4}(u)\dot N_{p}(u)=2L+64L M^{2}(u)N_{p}(u) $, the inner AH  (\ref{chichi}) is null while the nature of the  outer AH
(\ref{michi}) depends on the total value of the remaining terms in (\ref{naturep+}).

The Hawking temperature on the inner and outer AHs (\ref{chichi})  (\ref{michi})
of black hole can be determined using the surface gravity as
\begin{eqnarray}
\mathcal{T}_{AH^{-}}&=& \frac{1}{2\pi}\left(\frac{M(u)}{\left( 2M(u)+16M^{4}(u)N_{p}(u) +\mathcal{O}\left(N_{p}^2(u)\right) \right)^2} -N_p(u)\left( 2M(u)+16M^{4}(u)N_{p}(u) +\mathcal{O}\left(N_{p}^2(u)\right) \right)^2  \right)\nonumber\\
&=&\frac{1}{2\pi}\left(\frac{1}{4}\frac{1}{M(u)}-10 M^2(u)N_p(u) \right)+\mathcal{O}\left(N^2_{p}(u)\right),
\end{eqnarray}
and
\begin{eqnarray}
\mathcal{T}_{AH^{+}}&=& \frac{1}{2\pi}\frac{M(u)}{\left( \frac{1}{ N^\frac{1}{3}_{c}(u)}-\frac{2}{3}M(u)-\frac{8}{9}{M}^{2}(u) N^{\frac{1}{3}}_{p}(u)
-\frac{160}{81} M^3(u) N^{\frac{2}{3}}_p(u)-\frac{16}{3}M^4(u) N_p(u)+\mathcal{O}\left(N_{p}^\frac{4}{3}(u)\right)\right)^2}\nonumber\\
&&-\frac{3}{2\pi}N_c(u)\left( \frac{1}{ N^\frac{1}{3}_{c}(u)}-\frac{2}{3}M(u)-\frac{8}{9}{M}^{2}(u) N^{\frac{1}{3}}_{p}(u) -\frac{160}{81} M^3(u) N^{\frac{2}{3}}_p(u)-\frac{16}{3}M^4(u) N_p(u)+\mathcal{O}\left(N_{p}^\frac{4}{3}(u)\right) \right)^2 \nonumber\\
&=& \frac{1}{2\pi}\left(-\frac{3}{2}N^{\frac{1}{3}}_p(u) +3M(u)N^{\frac{2}{3}}_p(u)
+\frac{10}{3}M^2(u)N_p(u) )\right)+\mathcal{O}\left(N^{\frac{4}{3}}_{p}(u)\right).
\end{eqnarray}
respectively. One realizes that the temperature of the inner AH (\ref{chichi}) is always
positive while the temperature of the outer AH (\ref{michi}) is negative.
For the inner AH (\ref{chichi}), the background phantom
field decreases the temperature of the AH relative to the case of black hole in empty space. Here, regarding
(\ref{surp}), the negative temperature of the outer AH (\ref{michi}) is resulting from the fact that the phantom background
field effect dominates at large distances which leads to the negative surface gravity, namely the
gravitational repulsion.

Also, the semiclassical Bekenstein-Hawking entropy at the AHs (\ref{chichi})
 and  (\ref{michi}) are given by
\begin{equation}\label{sahp-}
\mathcal{S}_{AH^{-}}=\pi r_{AH^{-}}^{2}= \pi \left(4M^2(u) +64M^5(u) N_{p}(u)\right)
+\mathcal{O}\left(N^2_{p}(u)\right),
\end{equation}
and 
\begin{eqnarray}\label{sahp+}
\mathcal{S}_{AH^{+}}&=&\pi r_{AH^{+}}^{2}\nonumber\\
&=&\pi\left(\frac{1}{N^{\frac{2}{3}}_p(u)}-\frac{4}{3}\,{\frac{M(u)}{ N^{\frac{1}{3}}_p(u)}}
-\frac{4}{3} M^2(u) -\frac{224}{81} M^3(u) N^{\frac{1}{3}}_p(u)-\frac{1760}{243}M^4(u) N^{\frac{2}{3}}_p(u)-\frac{64}{3} M^5(u) N_p(u)\right)\nonumber\\
&&+\mathcal{O}\left(N_{p}^{\frac{4}{3}}(u)\right).
\end{eqnarray}
Then, for the phantom background with $N_p(u)\ll1$,  the inner AH (\ref{chichi})  has a negligible entropy in comparison  to the entropy associated with the outer AH (\ref{michi}). Also regrading (\ref{sahp-}), we see that the background phantom field contributes positively in the
entropy of the inner AH (\ref{chichi}), which leads  to a higher entropy relative to the single AH (\ref{ahempty}) of
the black hole in an empty background.
In the limit of $N_p(u)\rightarrow 0$, the entropy (\ref{sahp-}) of the inner AH (\ref{chichi})
tends to the entropy (\ref{sempty}) of AH (\ref{ahempty}) of
the black hole in an empty background while the entropy of the outer AH (\ref{michi}) blows up to infinity.
Also regarding (\ref{sahp+}), one realizes that since the entropy is a non-negative quantity, then this relation requires that the first term overcomes to all of the next terms. This requires the value of phantom field normalizing parameter $N_p(u)$ to be very small.
In fact, this entropy consideration on the cosmological AH (\ref{michi}) gives a restriction on the phantom field normalization parameter $N_p(u)$.

\section{Conclusion}
In this work, we have studied the thermodynamical features and dynamical evolutions of the various apparent horizons
associated with the Vaidya evaporating black hole surrounded by the cosmological fields of dust, radiation,
quintessence, cosmological constant-like and phantom. We have explored in detail how do these surrounding fields contribute to the characteristic features of a surrounded dynamical black hole. To this aim,  using the null vector decomposition of the spacetime metric and following
 York \cite{ahyork1}, we  have obtained the geometric properties of the solution in its general form. We have found that the solution to $g(\partial_u, \partial_u)=g_{uu}= 0$ representing
the TLS is the same as the solution to $\Theta=0$ representing
the AH location, except  the trivial solution $r=\infty$, for the latter. Then, the AH and the TLS  coincide for our spacetime (\ref{Vaidya}). Also, for the surrounded evaporating
black hole solutions here, we have verified that the condition $\frac{d\Theta}{du} \simeq0$ determining the location of EH is equal to the vanishing acceleration of the congruences of null geodesics at EH.
In the following, we summarize some of our results for the AH properties of
the evaporating black
hole in various backgrounds.  
\begin{itemize}
\item For the black hole with surrounding dust
field, the dust background appears with a positive surface
gravity increasing the gravitational attraction of the black hole. For this
case, the size of AH  is larger than   that of black hole in an
empty space.  For the black hole surrounded by a static  dust
field, i.e $\dot N_d (u)=0$,  AH is timelike surface and its dynamics is the same as the dynamics of AH in black hole in the empty space in which it is shrinking during the evaporation.
For a dynamical background with $\dot N_d (u)=2L$, the AH is a frozen null
surface. For the background with $-2L+\dot N_{d}(u)>0$, AH is spacelike surface
and expands during the evaporation while  for $-2L+\dot N_{d}(u)<0$, it is a timelike surface and shrinks during the evaporation.
The  temperature of
AH  for the black hole in dust field is lower than the temperature of AH of the black hole in an empty space. Finally, the entropy associated with the AH in a dust field background is higher than that of the black hole in an empty space.
\item For the black hole with surrounding radiation 
field, the surrounding radiation field contributes as the effective charge-like term  for the Vaidya black hole. For this case, the radiation background appears with a negative surface
gravity decreasing the gravitational attraction of the black hole, exactly
the same as the charge effect in Reissner-Nordström black hole.
For the extremal case of $\Delta(u)=0$, there is only one single 
AH equal to half of the dynamical Schwarzschild radius which is a timelike and  zero temperature surface, 
and shrinks during the evaporation. The entropy of the extremal AH   is  a quarter of the entropy associated to the single AH of the black hole in an empty space. For $\Delta(u)<0$, we have no real solutions while for $\Delta(u)>0$, there are two real physical AHs.
For the latter case, the contribution of the  background radiation field leads
to ($i$) the existence of a small inner AH   and ($ii$) an smaller outer AH  in comparison to the case of the AH for the black
hole in an empty background. For  the static surrounding radiation field,
i.e
$\mathcal{\dot N}_r(u)=0$,  the inner AH is expanding while the outer AH is shrinking during the evaporation, faster than the case of black hole in an empty background. For $\mathcal{\dot N}_{r}(u)>0$,  the inner and outer AH are expanding  and shrinking, respectively,   but both  moving faster than  the AH of the black hole in the static radiation background. 
For $\mathcal{\dot N}_{r}(u)<0$, it is possible that the inner and outer AH  move again outward  and inward, respectively,   but both  will
be slower than the AH
of the black hole in the  radiation backgrounds with $\mathcal{\dot N}_r(u)=0$
and $\mathcal{\dot N}_{r}(u)>0$. In the particular case of $L\,\mathcal{N}_r(u)=
-\mathcal{\dot N}_r(u) M(u)$, the inner AH
 will be frozen while the outer AH will
move inward exactly the same as AH  of the black hole in an
empty background.  For $\mathcal{\dot N}_r(u)\geq0$ the inner and outer AH  are spacelike
and timelike respectively, while for $\mathcal{\dot N}_r(u)<0$, we see that
the nature of AH depends on the value of $\frac{ \mathcal{\dot
N}_r(u)}{2M(u)}$. For the specific case of $L\,\mathcal{N}_r(u)=
-\mathcal{\dot N}_r(u) M(u)$, the inner AH will be null while the outer AH will be timelike. The temperature of the inner AH  is always negative while the temperature of the outer AH is positive, exactly the same as in the Reissner-Nordström
black hole. The inner AH   has a negligible entropy relative to the entropy associated with the outer AH in a dilute radiation background. The background radiation field contributes negatively in the
entropy of the outer AH leading to a lower entropy relative to the entropy
 of the single AH in an empty background.
\item For the black hole with surrounding quintessence 
field, the quintessence background contributes  negatively to the surface gravity which decreases the gravitational attraction of the black hole. This characteristic feature of the quintessence  field is favored by its cosmological   application as one of the candidates for
the dark energy.
For the extremal case of $\Delta(u)=0$, there is only one single  extremal
AH twice of the dynamical Schwarzschild radius which is a timelike and zero
temperature surface and shrinks during the black hole evaporation. 
The entropy of the extremal AH is four times the entropy
 associated to the single AH of the black hole in an empty space.
 For $\Delta(u)<0$, we have no real solutions while for $\Delta(u)>0$, there are two real physical AHs.
For the latter case, the contribution of the background quintessence field leads
to ($i$) the existence of an inner AH larger than the dynamical Schwarzschild radius $r(u)=2M(u)$, and ($ii$) a large outer cosmological AH  tending to infinity for $N_q(u)\lll1$. For the static surrounding quintessence field, i.e $\dot N_{q}(u)=0$, the inner and outer AH have equal velocities but with opposite signs. Then, the inner AH is a timelike surface and shrinks while the outer
is an spacelike surface and expands
during the evaporation in which both of these
AH are moving faster than the case of the single AH for the black hole in an empty space. For $\dot N_{q}(u)<0$, the inner AH   is a timelike surface
and shrinks while the outer is an spacelike surface and expands in which these are faster than the cases in static quintessence background and in empty spaces. For $\dot N_{q}(u)>0$, the behaviour and nature of the AH depend on the $M(u)$, $L$, $N_q(u)$ and $\dot N_q(u)$ values.
For the specific case of $4M^{2}(u)\dot N_{q}(u)=2L+8LM(u)N_{q}(u)$, the
inner AH is a frozen null surface while the outer AH  is shrinking timelike
surface. The temperature of the inner AH   is always positive while
the temperature of the outer AH is negative. For the inner AH, the background quintessence field decreases
the  temperature of the  AH relative to the case of black hole in empty
space. Finally, the inner AH has a negligible entropy in comparison  to the entropy associated with the outer AH. Also, the background quintessence field contributes positively in the
entropy of the inner AH, which leads  to a higher entropy relative to the single AH of the black hole in an empty background.
\item  For the black hole with surrounding cosmological  
field, the cosmological  filed contributes negatively to the surface gravity decreasing the gravitational attraction of the black hole. This characteristic feature of the cosmological
field is evidenced by its application as the
most favored candidates for the
dark energy. For the extremal case of $\Delta(u)=0$, there is only one single  extremal
AH, $3/2$ times the dynamical Schwarzschild radius, which is a timelike and zero
temperature surface and shrinks during the black hole evaporation. 
The entropy of the extremal AH is $\frac{9}{4}$ times the entropy
 associated to the single AH of the black hole in an empty space.
For $\Delta(u)<0$, we have no real solutions while for $\Delta(u)>0$, there are two real physical AHs.
For the latter case, the contribution of the background cosmological  field leads
to ($i$) the existence of an inner AH larger than the dynamical Schwarzschild radius $r(u)=2M(u)$, and ($ii$) a large outer cosmological AH  tending to infinity for $N_c(u)\lll1$.
For the static surrounding cosmological field, i.e $\dot N_{c}(u)=0$,  the inner AH is a timelike surface and shrinks faster than the case of the single AH for the black hole in an empty space while the outer
is an spacelike surface and expands
during the evaporation. For $\dot N_{c}(u)<0$, the inner AH   is a timelike surface
and shrinks while the outer is an spacelike surface and expands in which these are faster than the cases in static cosmological background and in empty spaces. For $\dot N_{c}(u)>0$, the behaviour and nature of the AH depend on the $M(u)$, $L$, $N_c(u)$ and $\dot N_c(u)$ values.
For the specific case of $8M^{3}(u)\dot N_{c}(u)=2L+24L M^{2}(u)N_{c}(u)$, the
inner AH is a frozen null surface while the outer AH  is shrinking timelike
surface. The temperature of the inner AH   is always positive while
the temperature of the outer AH is negative. For the inner AH, the background cosmological field decreases
the  temperature of the  AH relative to the case of black hole in empty
space. Finally, the inner AH has a negligible entropy in comparison  to the entropy associated with the outer AH. Also, we find that the background cosmological field contributes positively in the
entropy of the inner AH, which leads  to a higher entropy relative to the single AH of the black hole in an empty background.

\item  For the black hole with surrounding phantom  
field, the phantom  filed contributes negatively to the surface gravity which
decreases the gravitational attraction of the black hole. This characteristic feature of the phantom
field is evidenced by its application as one of the candidates for the
dark energy. For the extremal case of $\Delta(u)=0$, there is only one single  extremal
AH, $4/3$ times the dynamical Schwarzschild radius, which is a timelike and zero
temperature surface and shrinks during the black hole evaporation. 
The entropy of the extremal AH is $\frac{16}{9}$ times the entropy
 associated to the single AH of the black hole in an empty space.
For $\Delta(u)<0$, we have no real solutions while for $\Delta(u)>0$, there are two real physical AHs.
For the latter case, the contribution of the background phantom  field leads
to ($i$) the existence of an inner AH larger than the dynamical Schwarzschild radius $r(u)=2M(u)$, and ($ii$) a large outer cosmological AH  tending to infinity for $N_p(u)\lll1$.
For the static surrounding phantom field, i.e $\dot N_{p}(u)=0$,  the inner AH is a timelike surface and shrinks faster than the case of the single AH for the black hole in an empty space while the outer
is an spacelike surface and expands
during the evaporation. For $\dot N_{p}(u)<0$, the inner AH   is a timelike surface
and shrinks while the outer is an spacelike surface and expands in which these are faster than the cases in static cosmological background and in empty spaces. For $\dot N_{p}(u)>0$, the behaviour and nature of the AHs depend on the $M(u)$, $L$, $N_p(u)$ and $\dot N_p(u)$ values.
For the specific case of $16M^{4}(u)\dot N_{p}(u)=2L+64L M^{2}(u)N_{p}(u)$, the
inner AH is a frozen null surface while the outer AH  is shrinking timelike
surface. The temperature of the inner AH   is always positive while
the temperature of the outer AH is negative. For the inner AH, the background phantom field decreases
the  temperature of the  AH relative to the case of black hole in empty
space. Finally, the inner AH has a negligible entropy in comparison  to the entropy associated with the outer AH. Also, we find that the background phantom field contributes positively in the
entropy of the inner AH, which leads  to a higher entropy relative to the single AH of the black hole in an empty background.
\end{itemize}
 We aim to report elsewhere the detailed study of  EH properties together
with the back-reaction and metric fluctuation  for the surrounded Vaidya black holes. 


\end{document}